# Direct imaging of lattice strain-induced stripe phases in an optimally-doped manganite

L. Sudheendra, V. Moshnyaga, B. Damaschke and K. Samwer

*1.Physikalisches Institut, Friedrich-Hund-Platz 1, Universitaet Goettingen, D37077 Goettingen, Germany*

In a manganite film without quenched disorder, we show texturing in the form of insulating and metallic stripes above and below Curie temperature ($T_c$), respectively, by high resolution scanning tunneling microscopy/spectroscopy (STM/STS). The formation of these stripes involves competing orbital and charge orders, and are an outcome of overlapping electron wave-functions mediated by long-range lattice strain. Contrary to popular perception, electronically homogeneous stripe phase underlines the efficacy of the lattice strain in bringing about charge density modulation and in impeding the cross-talk between the order parameters, which otherwise evolves inhomogeneously in the form of orbitally-ordered insulating and orbitally disordered metallic phases.

Complex phases in manganites are a consequence of a direct interplay between spin, charge, and orbital order parameters [1, 2]. The resulting modulated phases observed in manganites analogous to other oxide [3] materials are termed electronically soft [4] as the competing order parameters show indifference towards the formation of either a homogeneous ordered or an inhomogeneous disordered phase. In colossal magnetoresistant (CMR) manganites [5], of the form $La_{1-x}Ca_xMnO_3$ ($0.2 \leq x \leq 0.33$), the localizing effect of the charge due to the Coulomb repulsion and Jahn-Teller (JT) [6,7] coupling at the Mn-site results in correlated polarons, with CE-type ordering, confined to nano-regions above $T_C$ [8-10]. Such phases symbolize the coupling of the spin, charge and the orbital order parameters brought about by the lattice degree of freedom. The increase of the one-electron bandwidth below $T_c$ results in a metallic state. The metallicity relaxes the condition for orbital and charge ordering, corroborated by the loss of correlation in the charge order parameter [8-10].

Apart from JT distortions, the transport properties of the manganite are strongly influenced by the size mismatch between the A-site cations (eg. La and Ca). The random Coulomb potential emanating from the quenched disorder at the A-site bears a strong influence on the electronic characteristics of manganites [11]. Indeed, electronic inhomogeneity, in the form of JT-distorted insulating and undistorted metallic phase, is justified based on the presence of such a disorder [12]. Further, as the JT distortions are stabilized by the additional rotational degrees of freedom [11], it is believed that the phase coexistence and the percolative nature of the metal-insulator transition (MIT) are also intrinsic to manganites [12, 13].

The absence of substrate-induced strain within the $La_{0.75}Ca_{0.25}MnO_3$ (LCMO) film, and in contrast, the presence of La-Ca cation-ordered rhombohedral superstructure was recently shown to suppress the electronic inhomogeneity down to 1 nm scale [14]. Nevertheless, the unit-cell deformation and the associated strain arising from the difference in the radii of the La and Ca cations is intrinsic and could persist even in the absence of A-site disorder. As the present study shows, homogeneous novel electronic phases can surface as a corollary to a co-operative effect involving distorted octahedra and the strain field generated by the ordering of the deformed unit-cells.

Epitaxial rhombohedral film exhibiting La and Ca ordering was grown by metal-organic aerosol technique [14]. STM and STS were performed at a base pressure of ~1-5×10$^{-10}$ torr. The film was loaded into the vacuum chamber after they were cleaned with *iso*-propanol solution. Mechanically cut Pt-Ir tips were employed to get high resolution images in the constant current mode (I = 0.1-0.3 nA) with a tip bias ranging from 0.35-0.7 V. Stable high resolution images were obtained on flat terraces of the as prepared film both at room and low temperatures (see supplementary information). Layer-by-layer growth with unit-cell heights of 4 Å and in certain cases integral multiples of 2 Å were detected by STM. The high resolution room temperature STM images on a 45×45 Å$^2$ area (Figs. 1a and b) demonstrate the stripe-like features with a periodicity of $\sqrt{2}a_p$ (where $a_p$ is the cubic-perovskite cell length). These stripes reveal two different corrugations: a large (black arrows in Fig. 1c) corrugation with a width (spread) of 6 Å, quite in contrast to the smaller corrugation (orange arrows in Fig. 1c), which is 4 Å wide. The distances between the two nearest large features are ~11 Å (Fig. 1b). High resolution STM at 115 K is visualized in Figs. 2a and b. One can clearly see the stripe-like features even in the ferromagnetic metallic state ($T_c \approx T_{MI}$ = 273 K). The peak to peak modulations are ~6.5-7.5 Å (~ $2a_p$). From Figs. 2a, and c, non-planar pyramidal-like structures of stripes are implicit. Such stripe features are contrary to expectations, as in the metallic state, a delocalization of $e_g$ electron quenches JT distorted octahedra ensuing screening of the charge.

Since stripes in an optimally hole-doped manganite are intriguing, we try to address the results obtained in the following order: Firstly, the origin of the stripe phase is elucidated based on lattice deformation and interrelated strain that manipulates the charge modulation and distortions at the Mn-site. In addition, the role of the strain in determining the coupling between the orbital and charge order parameters, and the intrinsically related transport property (insulating or metallic) of these stripes are clarified.

The criteria for obtaining atomic resolution by tunneling experiments in manganites was recently expounded, where the results point towards reduced screening of the charge due to defect-induced confinement – a trapped polaron - as one such possibility [15]. In hole-doped manganites, it is well known from x-ray scattering experiments that the number of such quasiparticles detectable, in general, is very small and it is almost non-existent in the rhombohedral phase [16]. This further augments the difficulties in obtaining atomic resolution by tunneling experiments on manganites [15, 17, 18].

Although the presence of small or large polarons can explain the near atomic resolution perceived, the formation of stripes with two different periodicities signify a different origin for the charge localization and structural modulation. Apart from point defect [15] and strong JT induced charge localization [17, 18], strain effects [19] especially originating from domain walls [4, 20] have great proclivity towards charge localization, and are also illustrated to yield exotic new phases in manganites. Atomic scale theory concerning lattice deformation, indeed, postulates such coupled electronic and elastic textures [21]. In the LCMO film, the presence of strain, at a qualitative level, can be described based on the atomic displacements caused by the ordering of the La-Ca cations. A checkered-board arrangement of uniaxially dilated La and compressed Ca unit-cells, can be deciphered from the cross-sectional high resolution transmission electron microscopy images (HRTEM) [14]. The unique ordering of uniaxially strained ($e_1$) unit-cells generates shear ($e_2$) and 'shuffle' ($s_x$ and $s_y$) deformations (Fig. 3a) [19]. These deformations, although symmetrically distributed, control the microscopic electronic and topological features of the manganite film [19]. Hence, the strain associated with the deformations effect structural distortions at the Mn octahedra; this in turn creates charge-disparity (Mn1 and Mn3) enhancing the atomic contrast in the tunneling experiments (Figs. 3b and 3c). As the ordering of the A-cation is long-range, these stripes point towards a correlated phenomenon relating charge modulation, structural distortions, and the associated orbital ordering. Such a correlation would be absent for small or large polaron description [16].

In manganites, based on the electron-phonon coupling and the long-range strain interactions, the presence of bond (along Mn-O-Mn directions or <100> directions in cubic notations) and diagonal (<110>) orbital ordering were theoretically predicted [22]. Such bond or diagonal-type stripes materialize when the inequality $C_{11}$-$C_{12}$-$2C_{44}$, where $C_{ij}$'s are elastic moduli, is either greater than zero or less than zero, respectively [22]. In line with this theory, as the structural distortions reflects in the orbital ordering, a periodicity of $\sqrt{2}a_p$ seen in Figs. 1a and 1b can be classified as diagonal orbital-ordered stripes. Although, there is an absence of a clear atomic lattice conceivably due to charge fluctuations. The CE-type stripes with two different LDOS features transpire due to the polarization of the electron wave-function involving next nearest neighbouring interactions (Fig. 3b) [22]. Similarly, the electron polarization along the Mn-O-Mn direction along *c* yields bond stripes (Fig. 3c). Recent Brillion scattering experiment on a similar composition has shown the increase in the $c_{11}$ with decrease in temperature [23]. Clearly, the shear and uniaxial deformations play a vital part in the stripe formation. It must be noted that the orbital stripes are predicted without the overlapping strain field from the lattice [22]. However, in reality, the structural distortions (JT distortions) are stabilized and the JT-induced strain is annealed out due to rotations of the octahedra.

The strain mediated coupling of the charge and the orbital orders introduces lattice "incommensuration" resulting in a CE-type of charge-orbital ordering, as can be comprehended from diagonal stripes (Fig. 3b). On the other hand, below $T_c$, a $2a_p$ periodicity of the bond stripes implies that the charge-orbital modulation is "commensurate" with the lattice illustrating a weakened charge-orbital coupling. Further, as Mn octahedra corresponding to La unit-cells (Fig. 3a) are associated with larger anisotropic distortion than the octahedra of La-Ca unit-cells, the similarly distorted octahedra, hence, have a periodicity of $2a_p$ rather than $a_p$ (Fig. 3c). It must be mentioned that both strain and magnetic interactions affect the bandwidth across the MIT and as the energetics (~ 20 meV/Mn) are similar [24, 25], these stripes provide signatures of possible magnetic interaction. Thus, such incommensurate and commensurate stripes [1, 4] can also be labeled as antiferro- and ferro- quadrupolar stripes, respectively arising from charge-orbital density wave.

As these stripes, in the LCMO film, are an outcome of inadequate screening of the electron wave-function due to the overlapping long-range strain interaction, it suggests a possible self-organization of polarons. Of relevance to this discussion is the nano-scale polaron-polaron interaction deduced from x-rays in a layered manganite [26]. The STS offers a better understanding of the intricacy of competing charge and orbital orders in engineering the polaron-polaron correlation, as the difference in the screening of the charge-carrier reflects on the conductance of the stripe phase. The tunneling current-voltage characteristic (Fig. 4) on the bond stripes reveals a metallic behavior (red curve) distinct from the diagonal stripes, which appear insulator-like (green curve). Therefore, the room temperature density of state ($\propto dI/dV$) with a depletion ($dI/dV_{V \to 0} \approx 0$) near the Fermi level (inset of Fig. 4) substantiates electron-hole localization and orbital ordering within diagonal stripes [27, 28]. Furthermore, the linear part of the I-V curve, which mirrors the 'Drude' part to the conductivity [17], appears less than 0.2 V in the bond stripe. Hence, the onset of nonlinear I-V characteristics at lower voltages reveals a dominant polaronic nature of the conductance. In addition, the metallicity and bond stripe entails charge delocalization and orbital ordering. Such coexistence could be comprehended by evoking an electronic inhomogeneity in the form of orbitally-ordered insulating and orbitally-disordered metallic state within the stripes. However, the conductance spectra on the bond stripes do not support a gap ($dI/dV_{V \to 0} \approx 0$) scenario. The electronic inhomogeneity in the form of insulating and metallic regions can, therefore, be ruled out. In contrast, this puzzle can be understood as arising from decoupling of the charge and orbital order parameters. Wherein, the strain modulates the structural distortions pinning the orbital, while in the limit of weak electron-phonon coupling, the electron itinerancy could be due to reduced dimensionality, as the stripes constitute a form of charge-orbital density wave structure [29].

In conclusion, in the limit of weak electron-phonon coupling, the polaron-polaron interaction leading to stripe structure mediated by elastic continuum due to the lattice mismatch provides the rationale for stripe in manganites [30]. These strain-induced stripes also provide insight into the complex structure-property relationship brought out by the propensity of manganites towards new phases such as a tendency towards a fragile CE phase within a rhombohedral structure [31], and also a polaron liquid-like metallic state [29] in the absence of quenched disorder or strong JT distortions.

**Acknowledgments**

Authors are grateful to P. B. Littlewood for helpful suggestions. Authors also thank O. Shapoval and A. Belenchuk for the help rendered in sample preparation; U. Waghmare, S. A. Koester and K. Gehrke for discussions. The work was supported by the Deutsche Forschungsgemeinschaft via SFB 602, project A2. L.S. acknowledges Alexander von Humboldt foundation for the fellowship.Emails: L. S (Lsudhee@gwdg.de) and K. S (ksamwer@gwdg.de)

**References**

[1] C. H. Chen, & S.-W. Cheong, *Phys. Rev. Lett.* **76**, 4042 (1996).

[2] J. C.Loudon, N. D. Mathur, & P. Midgley, *Nature* **420**, 797 (2002).

[3] S. A.Kivelson, E. Fradkin, & V. J. Emery, *Nature,* **393**, 550 (1998).

[4] G. C.Milward, M. J. Calderón, & P. B. Littlewood, *Nature*, **433**, 607 (2005).

[5] R. von Helmolt, J.Wecker, B. Holzapfel, L. Schultz, & K. Samwer, *Phys. Rev. Lett.* **71**, 2331 (1993).

[6] A. J. Millis, *Nature* **392**, 147 (1998).

[7] R. Kilian, & G. Khaliullin, *Phys. Rev. B* **60**, 13458 (1999).

[8] C. P. Adams, J. W. Lynn, Y. M. Mukovskii, A. A. Arsenov, & Shulyatev, D. *Phys. Rev. Lett.* **85**, 3954 (2000).

[9] P. Dai, *et al. Phys. Rev. Lett.* **85,** 2553 (2000).

[10] J. M. Zuo, & J. Tao, *Phys. Rev. B* **63**, 060407(R), (2001).

[11] L. M. Rodriguez-Martinez, & J. P. Attfield, *Phys. Rev. B* **58**, 2426 (1998).

[12] E. Dagotto, *Nanoscale Phase Separation and Colossal Magnetoresistance* (Springer Series in Solid State Science Vol. 136, Springer, Berlin, Germany (2002)).

[13] M. Uehara, S. Mori, C. H. Chen, & S.-W. Cheong, *Nature* **399**, 560 (1999).

[14] V. Moshnyaga, *et al.* Preprint at http://www.arxiv.org/cond-mat/0512350 (2005).

[15] H. M. Rønnow, Ch. Renner, G. Aeppli, T. Kimura, & Y. Tokura, *Nature*, **440**, 1025 (2006).

[16] V. Kiryukhin, *et al. Phys. Rev. B* **70**, 214424 (2004).

[17] Ch. Renner, G. Aeppli, B.-G. Kim, Y-A. Soh, & S.-W. Cheong, *Nature* 416, 518 (2002).

[18] J. X. Ma, D. T. Gillaspie, E. W. Plummer, & J. Shen, *Phys. Rev. Lett.* **95**, 237210 (2005).

[19] K. H. Ahn, T. Lookman, & A. R. Bishop, *Nature*, **428**, 401 (2004).

[20] K. H. Ahn, T. Lookman, A. Saxena, & A. R. Bishop, *Phys. Rev. B* **68**, 0921101 (2003).

[21] K. H. Ahn, T. Lookman, A. Saxena, & A. R. Bishop, *Phys. Rev. B* **71**, 212102 (2005).

[22] D. I. Khomskii, & K. I. Kugel, *Phys. Rev. B* **67**, 134401 (2003).

[23] Md Motin Seikh, C. Narayana, L. Sudheendra, A. K. Sood and C. N. R. Rao, *J. Phys.: Condens. Matter*, **16**, 4381 (2004).

[24] M. J. Calderón, A. J. Millis, & K. H. Ahn, *Phys. Rev. B* **68**, 100410 (R) (2003).

[25] S. Yunoki, T. Hotta & E. Dagotto, *Phys. Rev. Lett.* **84**, 3714 (2000).

[26] B. J. Campbell, *et al. Phys. Rev. B* **65**, 014427 (2001).

[27] L. Brey, & P. B. Littlewood, *Phys. Rev. Lett.* **95**, 117205 (2005).

[28] T. Hotta, A. Feiguin, & E. Dagotto, *Phys. Rev. Lett.* **86**, 4922 (2001).

[29] N. Mannella, *et al. Nature*, **438**, 474 (2005).

[30] J. C. Loudon *et al. Phys. Rev. Lett.* **94**, 097202 (2005).

[31] H. Aliaga, *et al. Phys. Rev. B* **68**, 104405 (2003).

**Figure captions**

**Figure 1**. Room temperature (294 K) STM images of La-Ca ordered $La_{0.75}Ca_{0.25}MnO_3$ film. **a and b,** The images (45×45 Å$^2$) represent the stripes originating from charge-orbital ordering. **c,** Line profiles corresponding to **Figs. a** and **b**. There are two distinct Mn sites (pink arrows)- large that are broader in comparison to the smaller sites indicated by black and orange arrows. Peak-to-peak distance is $\sim \sqrt{2} a_p$, indicating that the stripe modulation is along <110> directions. The tip is held at a positive voltage with respect to the sample; the peak features indicates an 'electron-like' character coming from the occupied state.

**Figure 2**. Stripes observed on LCMO film by STM at T << $T_c$. **a,** 49.5×49.5 Å$^2$ (115 K). **b,** 50×50 Å$^2$ (115 K). The images show stripes of Mn-octahedra with a peak-to-peak

(black arrows) distance of $\sim 2a_p$. The modulations of these stripes in certain images are 'pyramidal-like' (see line profiles of **a**).

**Figure 3.** Schematic representation of unit-cell deformations of the LCMO film, and the probable nearest and next nearest interactions leading to stripe phases. **a,** The Mn displacements deduced from cross-sectional (*a-b* plane) HRTEM[14]. The unit-cell lattice deformations are classified as long ($e_1$ and $e_2$) and short range ($s_x$ and $s_y$)[19]. Corners of the polygon are occupied by the Mn ion (termed as Mn-site). Strain-free superstructure in the *b-a* plane is shown by the orange square. **b,** One of the possible attractive next nearest interaction is depicted. The black square is the unit-cell in the *c-a* plane; Mn1, Mn2 and Mn4 are representative $Mn^{3+}$-like sites, while Mn3 is $Mn^{4+}$-like. The blue block arrows the direction of next nearest attraction and a measure of the electron polarization directions. The polarization of the electron wave-functions generates the two different LDOS features $Mn^{3+}$-like (Mn1) and $Mn^{4+}$-like (Mn3) sites predominantly observed within diagonal stripes. **c,** The nearest neighbour interactions generating bond stripes. The modulations (brown and green)- the direction of the charge-orbital density wave, which also is a measure of variation of the structural distortion.

**Figure. 4.** Current-voltage (I-V) characteristics of the stripe phases. Insulating-like behavior at 294 K (green) and metallic feature at 115 K (50×50 Å$^2$, red) are obtained on diagonal (STM is shown at top-left corner) and bond stripes (the corresponding STM is displayed bottom-right corner), respectively. Curves are average of four data sets taken at the points (yellow) selected in the STM images. The room temperature $dI/dV$ spectrum

(top-right inset) shows a depletion of LDOS near the Fermi level (V=0). The broken vertical lines indicate the linear part of the I-V curve.

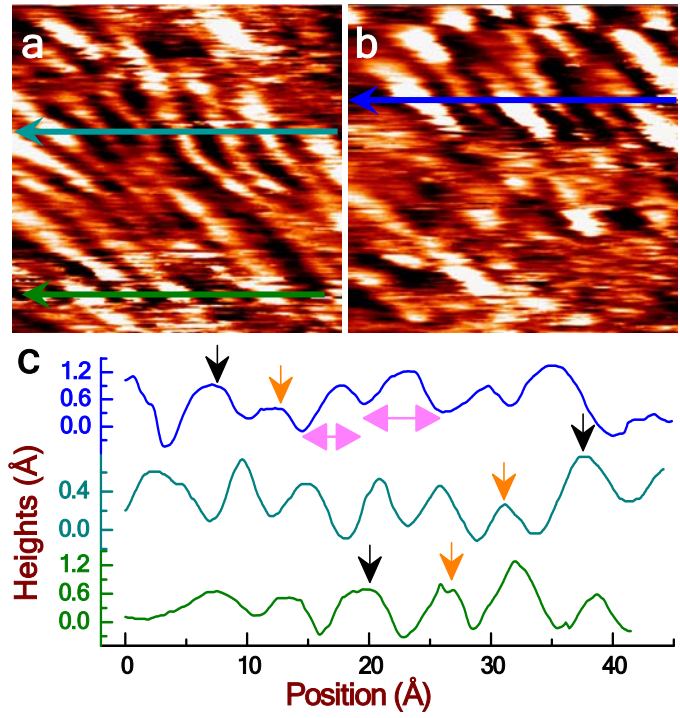

**Figure 1.**
**Sudheendra et al.**

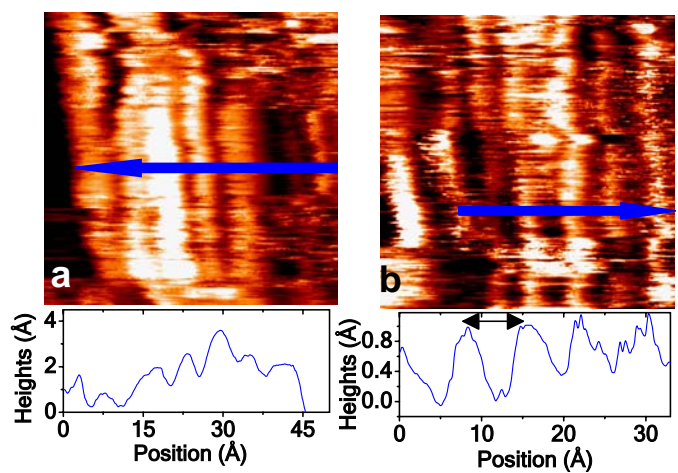

**Figure 2.
Sudheendra et al.**

**Figure 3.**
**Sudheendra et al.**

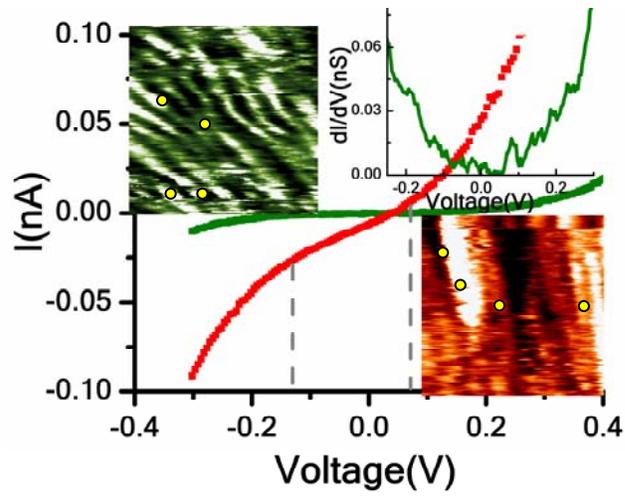

**Figure 4.
Sudheendra et al.**

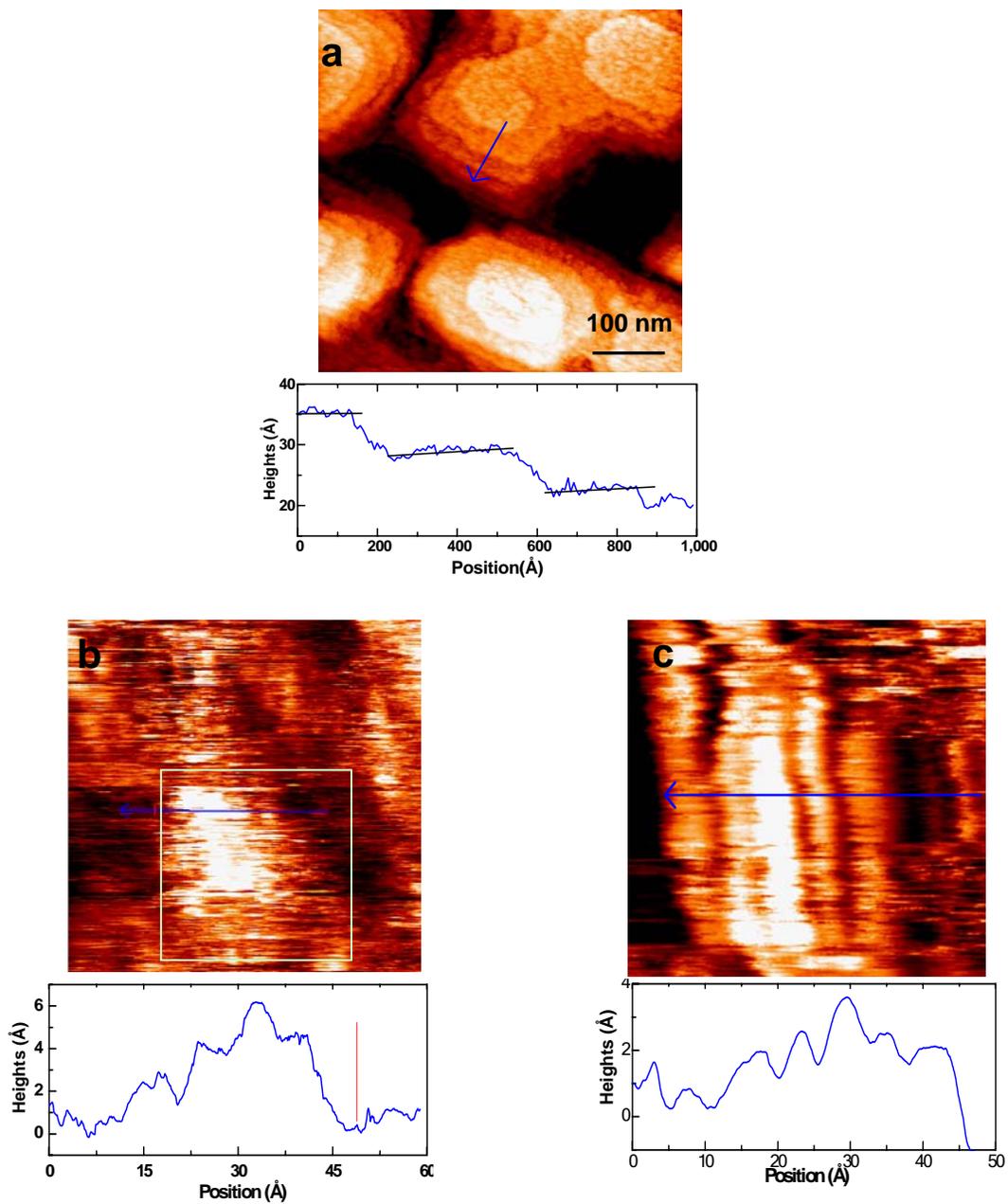

**Supplementary Figure 1**. Surface topography of the LCMO film obtained by STM (115 K). **a**, The layer-by-layer growth of the film is clearly visible. The growth of layers in the form of blocks indicates the strain-free growth. Widths of the terraces are around 2000 Å. The height of the layers is integral multiple of 4 Å. **b**, High resolution (100×100 Å$^2$) STM scanned on a terrace at 115 K. The vertical red line in the graph represents the position of the red mark in the view graph. **c**, Stripe phase obtained from scanning the square area (50×50 Å$^2$) marked in **b**.

17